\documentclass[preprint,preprintnumbers,showpacs,aps,prd,amssymb]{revtex4-1}

\usepackage{graphicx}
\usepackage{bm}
\usepackage{amsmath}
\usepackage{xcolor}

\def\calH{{\cal H}}

\def\calO{{\cal O}}

\def\calU{{\cal U}}

\def\Bbar{{\bar B}}

\def\hbar{{\bar h}}

\def\sbar{{\bar s}}

\def\SM{{\rm SM}}
\def\NP{{\rm NP}}
\def\GeV{{\rm GeV}}
\def\TeV{{\rm TeV}}
\def\Br{{\rm Br}}
\def\eff{{\rm eff}}
\def\LQ{{\rm LQ}}
\def\min{{\rm min}}
\def\dof{{\rm d.o.f.}}
\def\best{{\rm best}}

\def\nn{\nonumber}

\def\Ks{{K^{(*)}}}
\def\mhat{{\hat m}}
\def\shat{{\hat s}}
\def\uhat{{\hat u}}

\def\Bs2mumu{{B_s\to\mu^+\mu^-}}

\begin{document}
\title{Tree-level new physics in $R(\Ks)$}
\author{Jong-Phil Lee}
\email{jongphil7@gmail.com}
\affiliation{Sang-Huh College,
Konkuk University, Seoul 05029, Korea}

\begin{abstract}
We implement the $\chi^2$ fit for $R(\Ks)$ with possible tree-level new physics 
in a model-independent parametrization.
Relevant Wilson coefficients are decomposed into the new physics scale, its power, and the
fermionic couplings.
Constraints from the branching ratio of $B_s\to\mu^+\mu^-$ can be naturally incorporated 
with the scheme.
For a reasonable set of the parameter ranges it is found that the new physics is less than 
$\sim 5~\TeV$.
Some new physics models including the leptoquark, $Z'$, etc. can be embraced within our framework.
We give comments on new LHCb data which are close to the standard model predictions.
\end{abstract}

\maketitle
\section{Introduction}
The standard model (SM) of particle physics has been very successful for several decades 
and culminated in the discovery of the Higgs boson in 2012.
What is left is to go beyond the SM to find new physics (NP).
Flavor physics is one of the most promising field to observe NP.
One of the recent puzzle is related to the lepton universality violation (LUV) in the $b\to s$ transition 
where the ratio
\begin{equation}
R(\Ks)\equiv\frac{\Br(B\to\Ks\mu^+\mu^-)}{\Br(B\to\Ks e^+e^-)}~,
\end{equation}
shows a discrepancy between the experimental data and the SM predictions.
The process of $B\to\Ks\ell^+\ell^-$ is very interesting because it is a flavor-changing neutral current
and is not allowed at the tree level in the SM.
\par
Experimental results from the LHCb are \cite{LHCb2103,LHCb1705,Geng2103}
\begin{eqnarray}
R(K)[1.1,6.0] &=& 0.846^{+0.044}_{-0.041}~,  \nn\\
R(K^*)[0.045,1.1] &=& 0.660^{+0.110}_{-0.070}\pm0.024~, \nn\\
R(K^*)[1.1,6.0] &=& 0.685^{+0.113}_{-0.069}\pm0.047~,
\label{RKs_EXP}
\end{eqnarray}
where the square brackets are the bins of the momentum squared in $\GeV^2$.
On the other hand, the SM predictions for the $R(\Ks)$ are very close to unity \cite{Hiller0310,Bobeth0709,Geng1704}
\begin{eqnarray}
R(K)_\SM[1.0,6.0] &=&1.0004^{+0.0008}_{-0.0007}~, \nn\\
R(K^*)_\SM[0.045,1.1] &=& 0.920^{+0.007}_{-0.006}~, \nn\\
R(K^*)_\SM[1.1,6.0] &=& 0.996^{+0.002}_{-0.002}~.
\label{RKs_SM}
\end{eqnarray}
The apparent tension between experiments and the SM strongly suggests that NP would
play a role in the $R(\Ks)$ puzzle.
So far there have been many NP models to address the puzzle; 
supersymmetry (SUSY) \cite{Altmannshofer2002,Bardhan2107}, 
the leptoquark (LQ) \cite{Hiller1408,Dorsner1603,Bauer1511,Chen1703,Crivellin1703,Calibbi1709,Blanke1801,Nomura2104,Angelescu2103,Du2104,Cheung2204}, 
$Z'$ model \cite{Crivellin1501,Crivellin1503,Chiang1706,King1706,Chivukula1706,Cen2104,Davighi2105}, 
two-Higgs doublet model (2HDM) \cite{Hu1612,Crivellin1903,Rose1903}, 
unparticles \cite{JPL2106}, etc.
\par
Previously in \cite{JPL2110} we studied the $R(\Ks)$ puzzle in a different way.
For NP scenarios which contribute to $R(\Ks)$ at the tree level via the exchange of new
particles of mass $M_\NP$,
one can encapsulate their effects in new Wilson coefficients with parametrizations of
$C_{i\NP}\sim(v/M_\NP)^\alpha$, where $v$ is the SM vacuum expectation value.
Non-integer $\alpha$ might be possible for the unparticle-like scenarios, 
while SUSY, LQ, $Z'$, 2HDM, and so on correspond to $\alpha =2$.
In this context, Ref.\ \cite{JPL2110} could be called as model independent,
up to the form-factor dependence.
In \cite{JPL2110} it is assumed that $C_{9\NP}^\ell=-C_{10\NP}^\ell$ and
NP effects appear both in electron and muon sector.
In this framework constraints from $\Bs2mumu$ can be naturally included in the analysis.
\par
In this paper we adopt the framework of \cite{JPL2110} and turn off the NP effects in electron sector,
assuming that the lepton universality is violated {\em a priori.}.
Instead we allow $C_{10\NP}$ is independent of $C_{9\NP}$, 
and implement the $\chi^2$ fit for the relevant parameters 
making the analysis more quantitative than before.
\par
In the next Section the setup for our analysis is established, 
and Sec.\ III provides the results and discussions.
Conclusions appear in Sec.\ IV.
%
\section{Setup}
Let's first consider the effective Hamiltonian for the $b\to s\ell^+\ell^-$ transition,
\begin{equation}
\calH_{\rm eff} = 
-\frac{4G_F}{\sqrt{2}}V_{tb}V_{ts}^*\sum_i \left[
C_i(\mu)\calO_i(\mu)+C'_i(\mu)\calO'_i(\mu)\right]~.
\end{equation} 
The relevant operators for the $R(\Ks)$ anomaly are \cite{Geng2103,Alonso14,Geng17}
\begin{eqnarray}
\calO_9 &=& \frac{e^2}{16\pi^2}\left(\sbar\gamma^\mu P_L b\right)\left({\bar\ell}\gamma_\mu\ell\right)~,\nn\\
\calO_{10} &=& \frac{e^2}{16\pi^2}\left(\sbar\gamma^\mu P_L b\right)\left({\bar\ell}\gamma_\mu\gamma_5\ell\right)~.
\label{O9O10}
\end{eqnarray}
The primed operators are defined by replacing $P_L\to P_R$ in $\calO_{9,10}$,
which we do not consider in this analysis.
The matrix elements for $B\to\Ks$ can be expanded as \cite{Ali99}
\begin{eqnarray}
\langle K(p)|\sbar\gamma_\mu b|B(p_B)\rangle&=&
f_+\left[(p_B+p)_\mu-\frac{m_B^2-m_K^2}{s}q_\mu\right]+\frac{m_B^2-m_K^2}{s}f_0 q_\mu~,\\
\langle K(p)|\sbar\sigma_{\mu\nu} q^\nu(1+\gamma_5)b|B(p_B)\rangle&=&
i\left[(p_B+p)_\mu s -q_\mu(m_B^2-m_K^2)\right]\frac{f_T}{m_B+m_K}~,\\
\langle K^*(p)|(V-A)_\mu|B(p_B)\rangle&=&
-i\epsilon_\mu^*(m_B+m_{K^*})A_1+i(p_B+p)_\mu(\epsilon^*\cdot p_B)\frac{A_2}{m_B+m_{K^*}}\nn\\
&&
+iq_\mu(\epsilon^*\cdot p_B)\frac{2m_{K^*}}{s}(A_3-A_0)
+\frac{\epsilon_{\mu\nu\rho\sigma}\epsilon^{*\nu}p_B^\rho b^\sigma}{m_B+m_{K^*}}2V~,
\end{eqnarray}
where $f_{+,0,T}(s), ~A_{0,1,2}(s), ~T_{1,2,3}(s), V(s)$, and $f_-=(f_0-f_+)(1-\mhat_K^2)/\shat$ are the form factors.
Here,
\begin{eqnarray}
 q&=&p_B-p~,~~~s=q^2=(p_B-p)^2~,\\
 \shat&=&\frac{s}{m_B^2}~,~~~\mhat_i=\frac{m_i}{m_B}~.
 \end{eqnarray}
As for the form factors we adopt the exponential function \cite{Ali99}
\begin{equation}
F(\shat)=F(0)\exp\Big(c_1\shat+c_2\shat^2+c_3\shat^3\Big)~,
\label{expFF}
\end{equation}
where $c_i$ are the related coefficients.
\par
The decay rates for $B\to\Ks\ell^+\ell^-$ are obtained by integrating the corresponding
differential decay rates \cite{Chang2010}
\begin{eqnarray}
\frac{d\Gamma_K}{d\shat}&=&
\frac{G_F^2\alpha^2m_B^5}{2^{10}\pi^5}|V_{tb}V_{ts}^*|^2\uhat_{K,\ell}\left\{
(|A'|^2+|C'|^2)\left(\lambda_K-\frac{\uhat_{K,\ell}^2}{3}\right)
+|C'|^24\mhat_\ell^2(2+2\mhat_K^2-\shat) \right .\nn\\
&&\left. +{\rm Re}(C'D'^*)8\mhat_\ell^2(1-\mhat_K^2)+|D'|^24\mhat_\ell^2\shat\right\}~,
\\
\frac{d\Gamma_{K^*}}{d\shat}&=&
\frac{G_F^2\alpha^2m_B^5}{2^{10}\pi^5}|V_{tb}V_{ts}^*|^2\uhat_{K^*,\ell}\left\{
\frac{|A|^2}{3}\shat\lambda_{K^*}\left(1+\frac{2\mhat_\ell^2}{\shat}\right)
+|E|^2\shat\frac{\uhat_{K^*,\ell}^2}{3}\right.\nn\\
&&
+\frac{|B|^2}{4\mhat_{K^*}^2}\left[\lambda_{K^*}-\frac{\uhat_{K^*,\ell}t^2}{3}+8\mhat^2_{K^*}(\shat+2\mhat_\ell^2)\right]
+\frac{|F|^2}{4\mhat_{K^*}^2}\left[\lambda_{K^*}-\frac{\uhat_{K^*,\ell}^2}{3}+8\mhat_{K^*}^2(\shat-4\mhat_\ell^2)\right]\nn\\
&&
+\frac{\lambda_{K^*}|C|^2}{4\mhat_{K^*}^2}\left(\lambda_{K^*}-\frac{\uhat_{K^*,\ell}^2}{3}\right)
+\frac{\lambda|_{K^*}|G|^2}{4\mhat_{K^*}^2}\left[\lambda_{K^*}-\frac{\uhat_{K^*,\ell}^2}{3}
	+4\mhat_\ell^2(2+2\mhat_{K^*}^2-\shat)\right]\nn\\
&&
-\frac{{\rm Re}(BC^*)}{2\mhat_{K^*}^2}\left(\lambda_{K^*}-\frac{\uhat_{K^*,\ell}^2}{3}\right)(1-\mhat_{K^*}^2-\shat)\nn\\
&&
-\frac{{\rm Re}(FG^*)}{2\mhat_{K^*}^2}\left[\left(\lambda_{K^*}-\frac{\uhat_{K^*,\ell}^2}{3}\right)(1-\mhat_{K^*}^2
	-\shat)-4\mhat_\ell^2\lambda_{K^*}\right]\nn\\
&&\left.
-\frac{2\mhat_\ell^2}{\mhat_{K^*}^2}\lambda_{K^*}\left[{\rm Re}(FH^*)-{\rm Re}(GH^*)(1-\mhat_{K^*}^2)\right]
+\frac{\mhat_\ell^2}{\mhat_{K^*}^2}\shat\lambda_{K^*}|H|^2\right\}~,
\end{eqnarray}
where the kinematic variables are
\begin{eqnarray}
\lambda_H&=&1+\mhat_H^4+\shat^2-2\shat-2\mhat_H^2(1+\shat)~,\\
\uhat_{H,\ell}&=&\sqrt{\lambda_H\left(1-\frac{4\mhat_\ell^2}{\shat}\right)}~.
\end{eqnarray}
The auxiliary functions $A',\cdots, D'$ and $A,\cdots, H$ are defined by the Wilson coefficients and the form factors \cite{Ali99},
%
\begin{eqnarray}
A'&=&C_9 f_+ +\frac{2\mhat_b}{1+\mhat_K}C_7^\eff f_T ~,\\
B'&=&C_9 f_- -\frac{2\mhat_b}{\shat}(1-\mhat_K)C_7^\eff f_T ~,\\
C'&=&C_{10} f_+ ~,\\
D'&=&C_{10} f_- ~,
\end{eqnarray}
and
\begin{eqnarray}
A&=&\frac{2}{1+\mhat_{K^*}}C_9 V+\frac{4\mhat_b}{\shat}C_7^\eff T_1~,\\
B&=&(1+\mhat_{K^*})\left[C_9 A_1+\frac{2\mhat_b}{\shat}(1-\mhat_{K^*})C_7^\eff T_2\right]~,\\ 
C&=&\frac{1}{1-\mhat_{K^*}^2}\left[(1-\mhat_{K^*})C_9 A_2
	+2\mhat_b C_7^\eff \left(T_3+\frac{1-\mhat_{K^*}^2}{\shat}T_2\right)\right]~,\\
D&=&\frac{1}{\shat}\left\{C_9\left[(1+\mhat_{K^*})A_1-(1-\mhat_{K^*})A_2
	-2\mhat_{K^*}A_0\right]-2\mhat_b C_7^\eff T_3\right\}~,\\
E&=&\frac{2}{1+\mhat_{K^*}}C_{10} V ~,\\
F&=&(1+\mhat_{K^*})C_{10} A_1 ~,\\
G&=&\frac{1}{1+\mhat_{K^*}}C_{10} A_2 ~,\\
H&=&\frac{1}{\shat}C_{10}\left[(1+\mhat_{K^*})A_1-(1-\mhat_{K^*})A_2-2\mhat_{K^*} A_0\right] ~.
\end{eqnarray}
%
\par
Now we rewrite the Wilson coefficient as
\begin{eqnarray}
C_{9,10\NP}^\ell = A_{9,10}^\ell\left(\frac{v}{M_{\NP}}\right)^\alpha ~,
\label{C_setup}
\end{eqnarray}
where $\ell=e,\mu$.
Here $M_\NP$ is the NP scale and $A_{9,10}^\ell$ are the involved coefficients, and
$\alpha$ is a free parameter.
$A_{9,10}^\ell$ involve the fermionic couplings to NP.
In this analysis we simply put $A_{9,10}^e=0$.
With this setup we assume the NP effects only in the muon sector.
Since we do not allow NP in the electron sector, the LUV is assumed to exist 
{\em a priori} in this analysis.
If $A_{9,10}^e\ne 0$, they could affect the electron dipole moment (EDM) which fits very well with
the SM predictions.
It means that $A_{9,10}^e$ would be severely constrained by the EDM.
\par
%
%
%
It is possible that NP (SUSY, for example) could contribute to $C_7$.
But the constraint from $B\to X_s\gamma$ puts strong bounds on $\delta C_7$ as
\begin{equation}
-0.032 \le \delta C_7 \le 0.027~,
\end{equation}
at the $2\sigma$ level \cite{Bardhan2107}.
It seems that there is little room for NP to affect $C_7$ a lot,
and we do not consider the NP effects on $C_7$ later on.
\par
The setup of Eq.\ (\ref{C_setup}) is useful since one could consider the NP scale $M_\NP$ and 
the fermionic coupling part separately and model-independent way \cite{JPL2110}.
Table \ref{models} shows $A_9^\mu$ and $M_\NP$ for some models.
\begin{table}
\begin{tabular}{c|ccc}\hline
& SUSY \cite{Altmannshofer2002} & LQ \cite{Du2104} & $Z'$ \cite{Cen2104} \\\hline\hline
\rule{0pt}{25pt}
$A_9^\mu$ 
& \scalebox{1.5}
{$~~~-\frac{1}{16e^2}\frac{X_{bs}X_{\mu\mu}}{V_{tb}V^*_{ts}} ~~~$}
& \scalebox{1.5}{$~~~-\frac{\pi x_{s\mu} x_{b\mu}^*}{\alpha_{em}V_{tb}V_{ts}^*}~~~$}
& \scalebox{1.5}
{$~~~\frac{2\pi{\tilde g}^2 V_{L22}^d V_{L32}^{d*}}{3\alpha_{em}V_{tb}V_{ts}^*}~~~$} \\\hline
$M_\NP$ & sbottom mass & LQ mass & $Z'$ mass \\\hline
\end{tabular}
\caption{$A_9^\mu$ and $M_\NP$ for some models where $\alpha=2$.}
\label{models}
\end{table}
It should be noted that in Table \ref{models} $A_9^\mu$ of SUSY comes from the
sbottom loop, and other loop contributions involving sneutrino, stau, etc.
also exist \cite{Altmannshofer2002} which are not our concern here. 
In models of the Table \ref{models}, $\alpha$ is fixed to be $\alpha=2$.
For unparticles the new effects appear as $\sim(v^2/\Lambda_\calU^2)^{d_\calU}$, 
where $\Lambda_\calU\gtrsim 1~{\rm TeV}$ is the unparticle scale and 
$d_\calU$ is the scaling dimension of the unparticle operator.
The scale invariance of the unparticle makes it possible for $d_\calU$ to be non-integers \cite{Georgi}.
\par
Since $R(\Ks)$ involves the $b\to s\ell^+\ell^-$ transition, 
main constraints for the relevant parameters come from $B_s\to\mu^+\mu^-$ decay.
If NP effects are included the branching ratio of the $B_s\to\mu^+\mu^-$ can be written as
\begin{equation}
\Br(B_s\to\mu^+\mu^-)_\calU 
= \Br(B_s\to\mu^+\mu^-)_\SM \left|1
	+\frac{C_{10\NP}^\mu}{C_{10\SM}^\mu}\right|^2~,
\end{equation}
where $C_{10\SM}^\mu$ is the SM value of $C_{10}^\mu$ \cite{Damir1205}and
the SM prediction is \cite{Geng2103,Beneke1908}
\begin{equation}
\Br(\Bs2mumu)_\SM = (3.63\pm0.13)\times 10^{-9}~.
\label{BrBsmumuSM}
\end{equation}
%
%
%
On the other hand, the experimental data is \cite{HFLAV-ICHEP22}
\begin{equation}
\Br(B_s\to\mu^+\mu^-)_{\rm HFLAV} = (3.45\pm0.29)\times 10^{-9}~,
\label{BrBsmumuEXP}
\end{equation}
from the heavy flavor averaging group(HFLAV).
The average includes recent CMS measurements of the $\Bs2mumu$ decays \cite{CMS-006},
\begin{equation}
\Br(B_s\to\mu^+\mu^-)_{\rm CMS} = (3.83^{+0.38+0.24}_{-0.36-0.21})\times 10^{-9}~,
\end{equation}
so the averaged value gets closer to the SM value.
%
%
%
\section{Results and Discussions}
In this analysis our parameters span $-100\le A_{9,10}^\mu\le 100$, $0\le\alpha\le 5$, and
$1~\TeV\le M_\NP\le 10~\TeV$.
We implement the $\chi^2$ fit to the experimental data of $R(\Ks)$.
Our best-fit values are summarized in Table \ref{best_fit}.
\begin{table}
\begin{tabular}{ccccccc}\hline
$\alpha$ & $M_{\rm NP}$ (TeV) & $~~~A_9^\mu~~~$ & $~~~A_{10}^\mu~~~$  &
   $R(K)[1.1., 6.0]$ & $R(K^*)[0.045,1.1]$ & $R(K^*)[1.1, 6.0]$  \\ \hline\hline
$1.24$ & $2.54$ & $30.6$ & $89.3$ &
   $0.838$ & $0.805$ & $0.734$ \\\hline 
\end{tabular}
\caption{Best-fit values}
\label{best_fit}
\end{table}
The minimum value of $\chi^2$ ($\chi^2_{\min}$) per d.o.f. is $2.15$.
%
%
%
We find that $\chi^2_\min/\dof =2.15\simeq \chi^2_{p=0.116}/\dof$
where $\chi^2_p$ is the value of $\chi^2$ for which the probability of $\chi^2>\chi^2_p$ is $p$.
%
%
%
%
\par
\begin{figure}
\begin{tabular}{cc}
\hspace{-1cm}\includegraphics[scale=0.13]{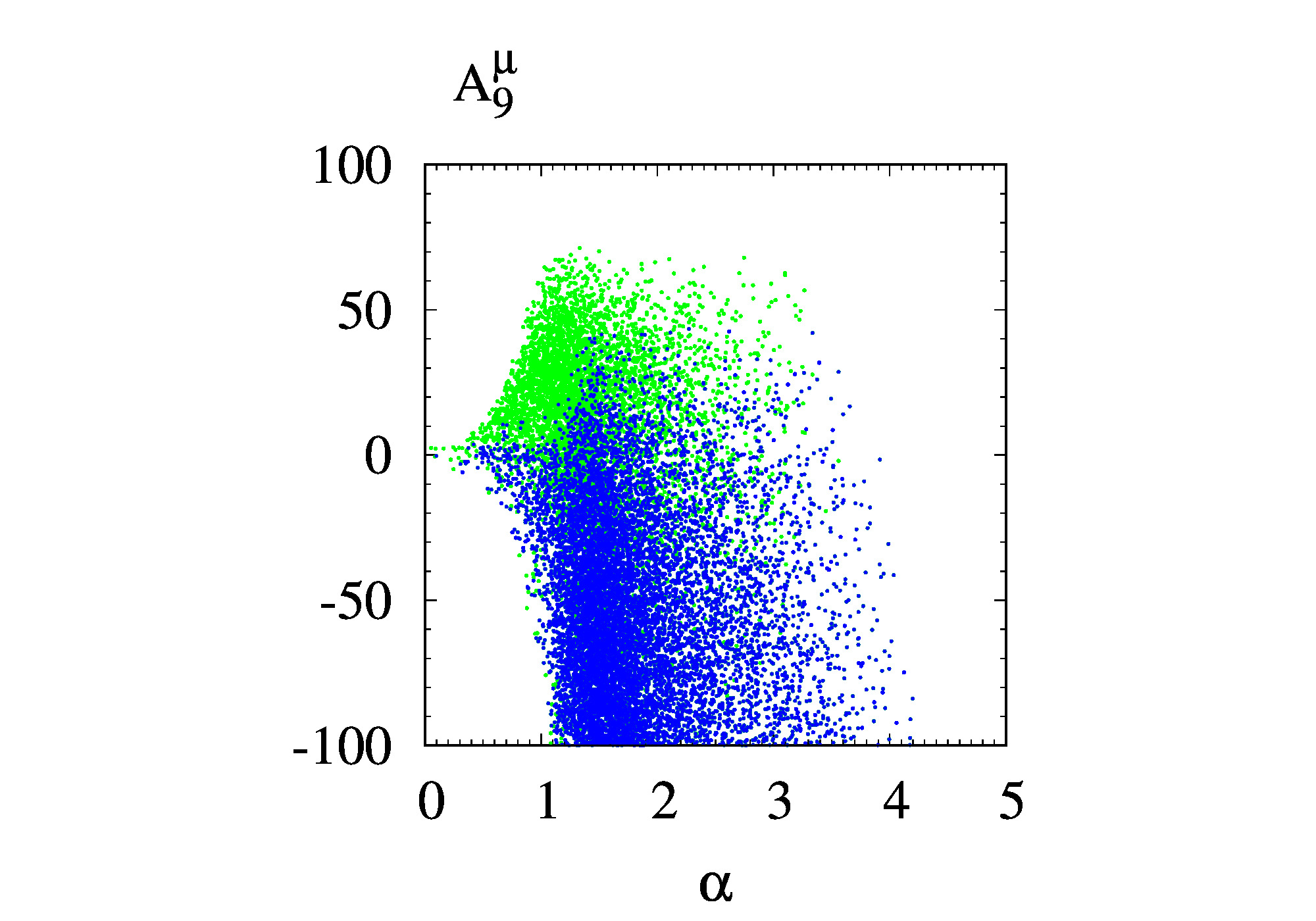}&
\hspace{-1cm}\includegraphics[scale=0.13]{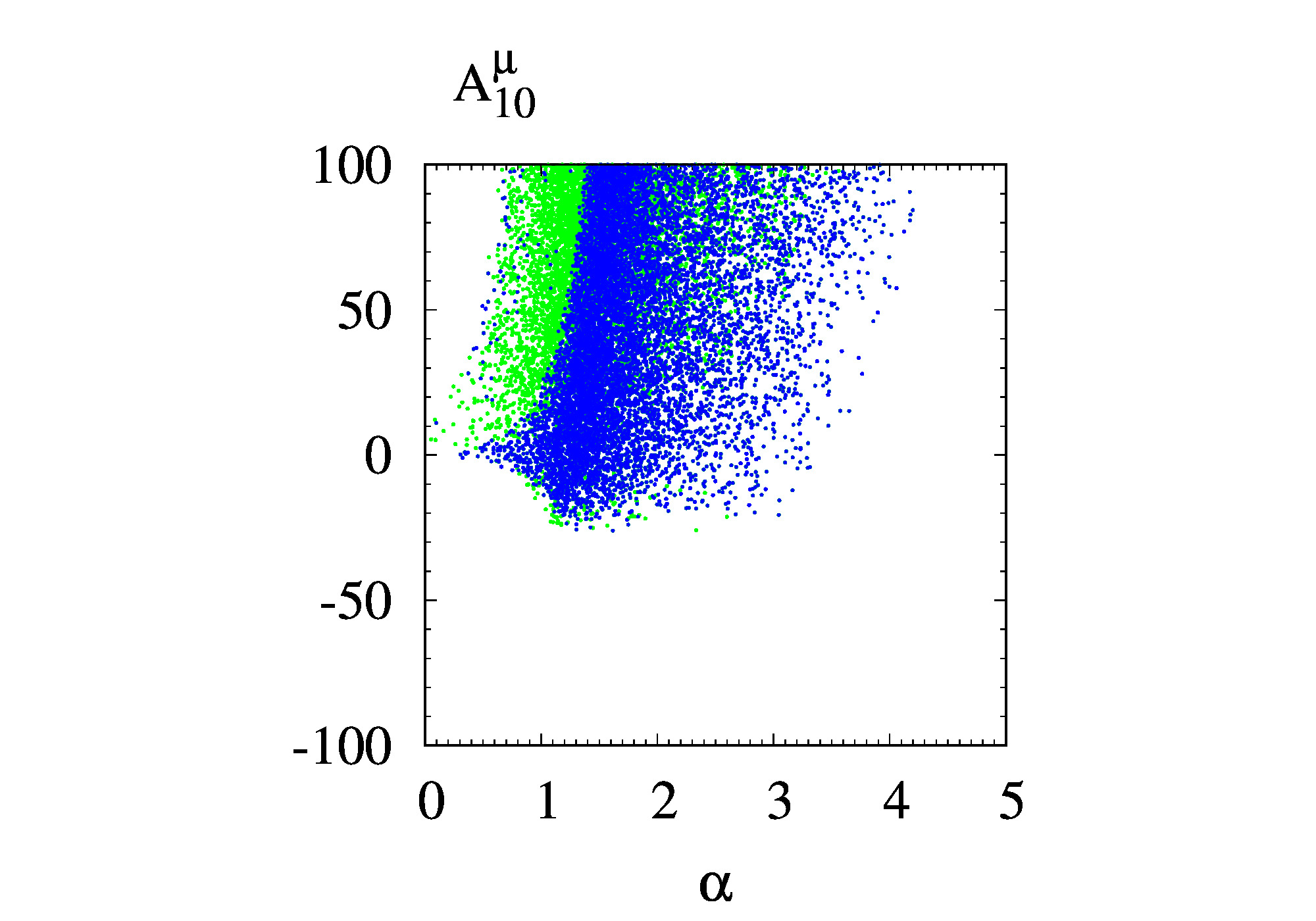}\\
\hspace{-1cm} (a) & \hspace{-1cm} (b)\\
\hspace{-1cm}\includegraphics[scale=0.13]{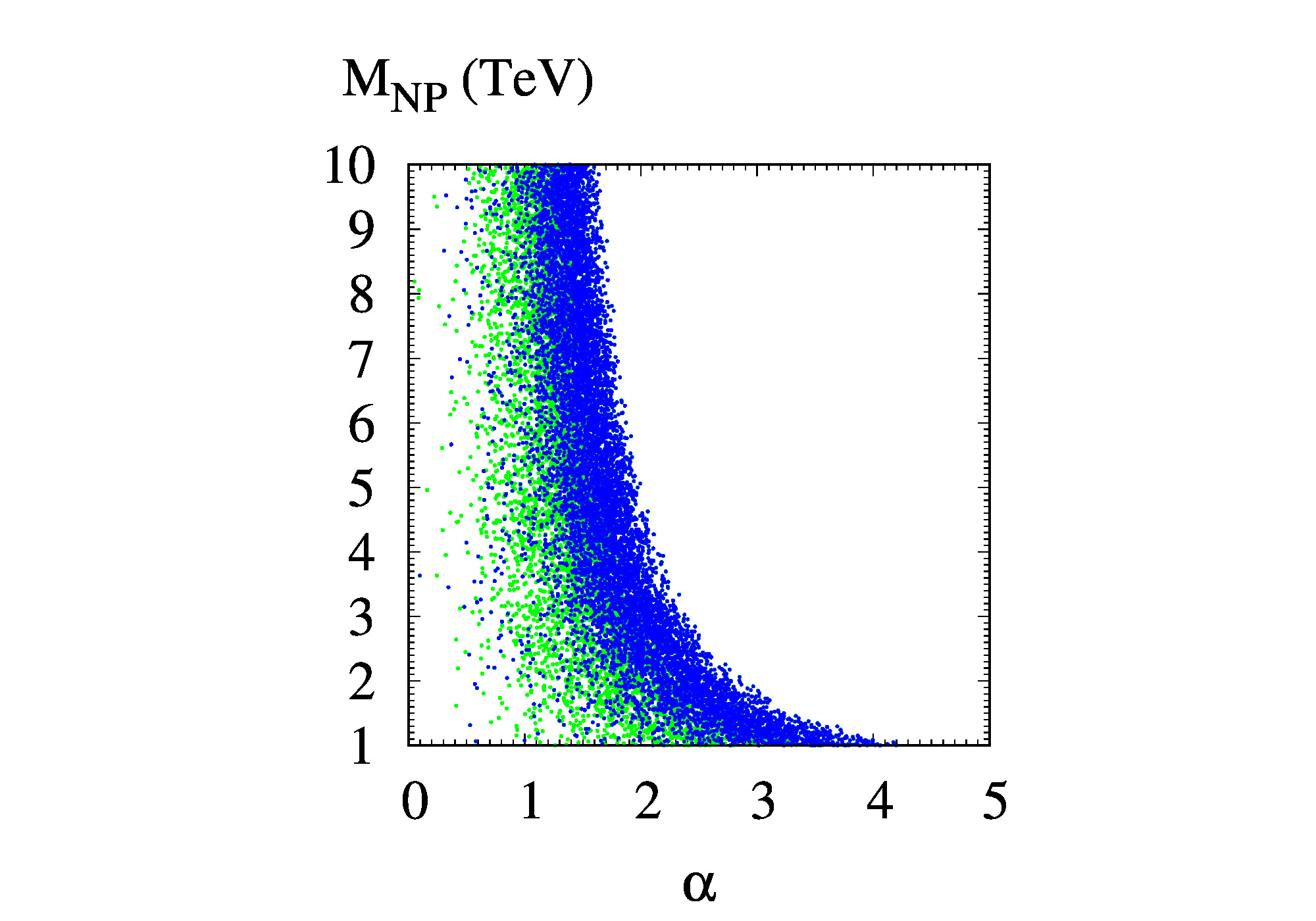} & \\
\hspace{-1cm} (c) &
\end{tabular}
\caption{\label{F1} Allowed regions of (a) $A_9^\mu$ vs $\alpha$, (b) $A_{10}^\mu$ vs $\alpha$,
and (c) $M_\NP$ vs $\alpha$ at the $2 \sigma$ level by $R(\Ks)$ (green).
Blue dots represent the subregion of the green ones that satisfy the $B_s\to\mu^+\mu^-$ constraint.
}
\end{figure}

Figure \ref{F1} shows the allowed regions of the relevant parameters.
Green dots represent the $2\sigma$ regions of the $\chi^2$ fit to $R(\Ks)$, 
and blues dots are the subregions that satisfy the $\Bs2mumu$ constraint.
As in Fig.\ \ref{F1}, large $|A_9^\mu|$ is not allowed for small $\alpha\lesssim 0.5$.
The region is forbidden because NP effects could be large there.
This is also almost true for blue dots of $A_{10}^\mu$ in Fig.\ \ref{F1} (b).
Figure \ref{F1} (c) depicts a very similar behavior to that of the corresponding parameters
in the unparticle scenario.
Maximum values of $M_\NP$ reduce greatly for $\alpha\gtrsim 2$.
In our parameter window of $|A_{9,10}^\mu|\le 100$ and $M_\NP\le 10~\TeV$ 
only small values of $\alpha\lesssim 4.2$ are allowed,
resulting in the best-fit value of $\alpha$ being $\alpha_{b.f.}=1.24$
where $b.f.$ means the best fit.
We find that for larger $\alpha\gtrsim 2$, 
the value of $\chi^2_\min$ for a given $\alpha$ gets significantly larger,
while $\chi^2_\min$ values are rather insensitive to $M_\NP$.
%
%
\par
\begin{figure}
\begin{tabular}{cc}
\hspace{-1cm}\includegraphics[scale=0.13]{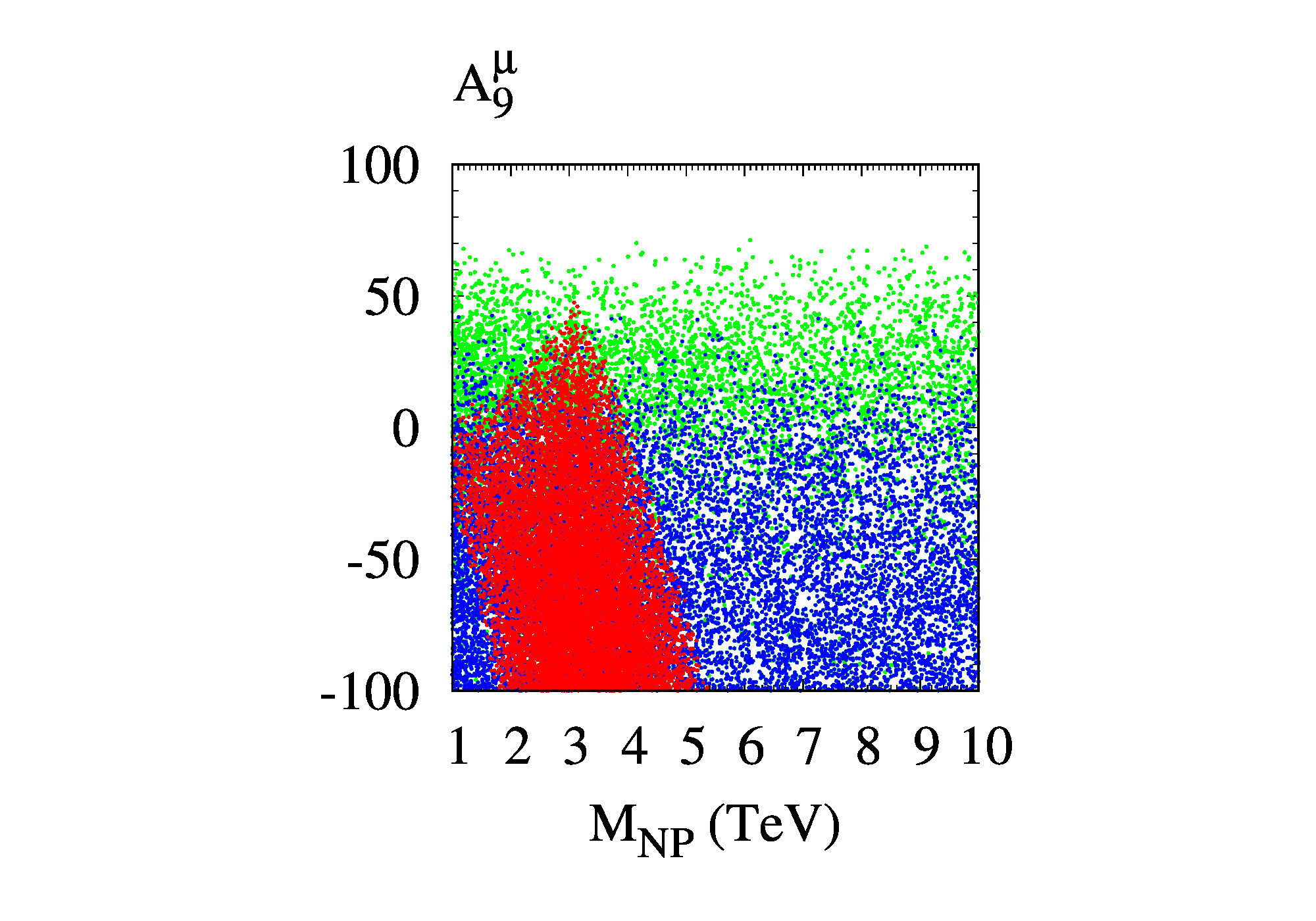}&
\hspace{-1cm}\includegraphics[scale=0.13]{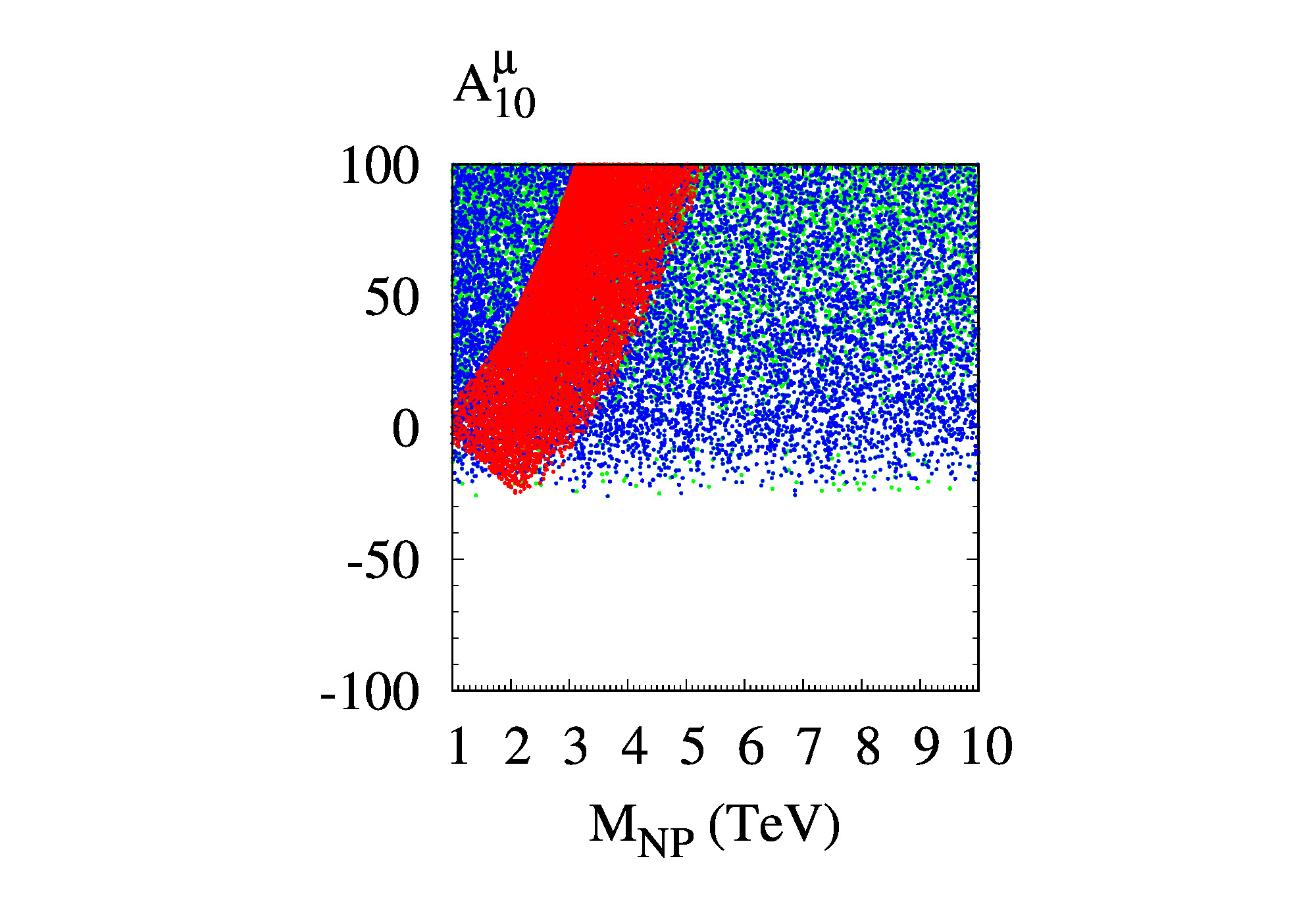} \\
\hspace{-1cm} (a) & \hspace{-1cm} (b) \\
\hspace{-1cm}\includegraphics[scale=0.13]{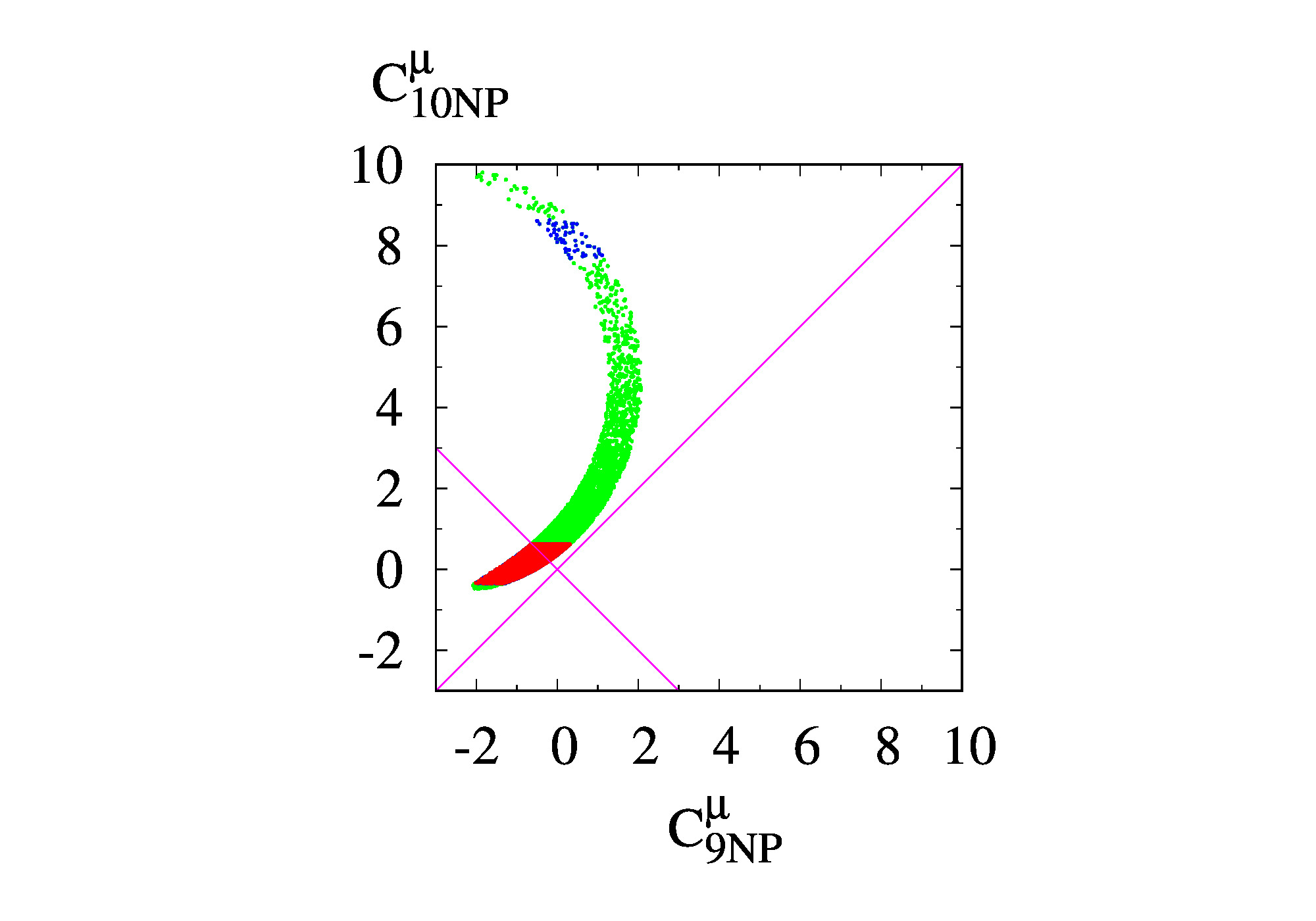} & \\
\hspace{-1cm} (c) &
\end{tabular}
\caption{\label{F2} Allowed regions of (a) $A_9^\mu$ vs $M_\NP$, (b) $A_{10}^\mu$ vs $M_\NP$,
and (c) $C_{10\NP}^\mu$ vs $C_{9\NP}^\mu$ at the $2\sigma$ level by $R(\Ks)$ (green).
Blue dots represent the subregion of the green ones that satisfy the $B_s\to\mu^+\mu^-$ constraint, and red dots for $\Bs2mumu$ and $\alpha=2$.
Magenta lines in (c) represent $C_{9\NP}^\mu=\pm C_{10\NP}^\mu$.}
\end{figure}

Figure \ref{F2} shows another aspects of the parameter space.
In Figs.\ \ref{F2} (a) and (b) allowed $A_{9,10}^\mu$ vs $M_\NP$ are shown. 
Here we specify the region where $\alpha=2$ with red dots.
Note that $M_\NP$ is allowed for $\lesssim 5$ TeV for fixed $\alpha=2$..
For larger $M_\NP$ NP effect gets smaller, which is not good to explain the data.
If we extend the parameter window wider, e.g., $|A_{9,10}^\mu|\ge 100$, 
then larger $M_\NP\gtrsim 5\TeV$ could be also possible .
This is because it is the Wilson coefficients $C_{9,10\NP}^\mu$ that directly affect the values of
$R(\Ks)$.
For given $C_{9,10\NP}^\mu$, large $|A_{9,10}^\mu|$ is compensated by large $M_\NP$.
%
\par
One can see in Fig.\ \ref{F2} (b) that when the constraint of $\Bs2mumu$ is imposed 
$A_{10}^\mu$ is mostly positive.
The reason is that $C_{10\NP}^\mu/C_{10\SM}^\mu$ favors negative values 
to fit the experimental data, and $C_{10\SM}^\mu<0$ \cite{Buras1303}.
The Wilson coefficients $C_{9,10\NP}^\mu$ is presented in Fig.\ \ref{F2} (c).
Compared to other Figures like Figs.\ \ref{F1} (a)-(c) and Figs.\ \ref{F2} (a)-(b), 
$C_{9,10\NP}^\mu$ are significantly constrained by $\Bs2mumu$. 
Figure \ref{F2} (c) also shows that $C_{9\NP}^\mu=-C_{10\NP}^\mu$ is still a good solution
to the $R(\Ks)$ puzzle.
It should be also noted that the SM ($C_{9\NP}^\mu=C_{10\NP}^\mu=0$) is slightly off 
of the allowed region at the $2\sigma$ level.
\par
\begin{figure}
\begin{tabular}{cc}
\hspace{-1cm}\includegraphics[scale=0.13]{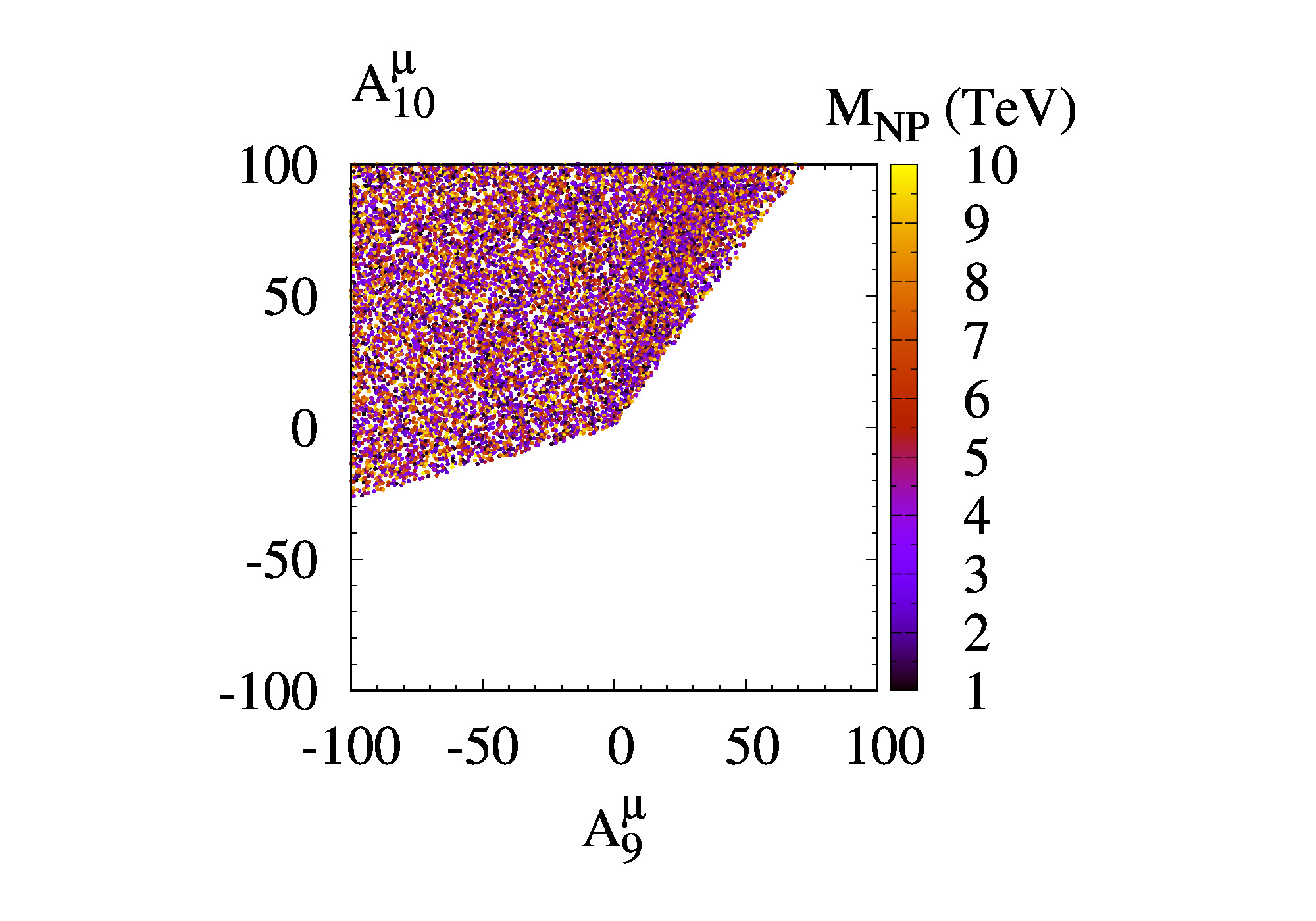} &
\hspace{-1cm}\includegraphics[scale=0.13]{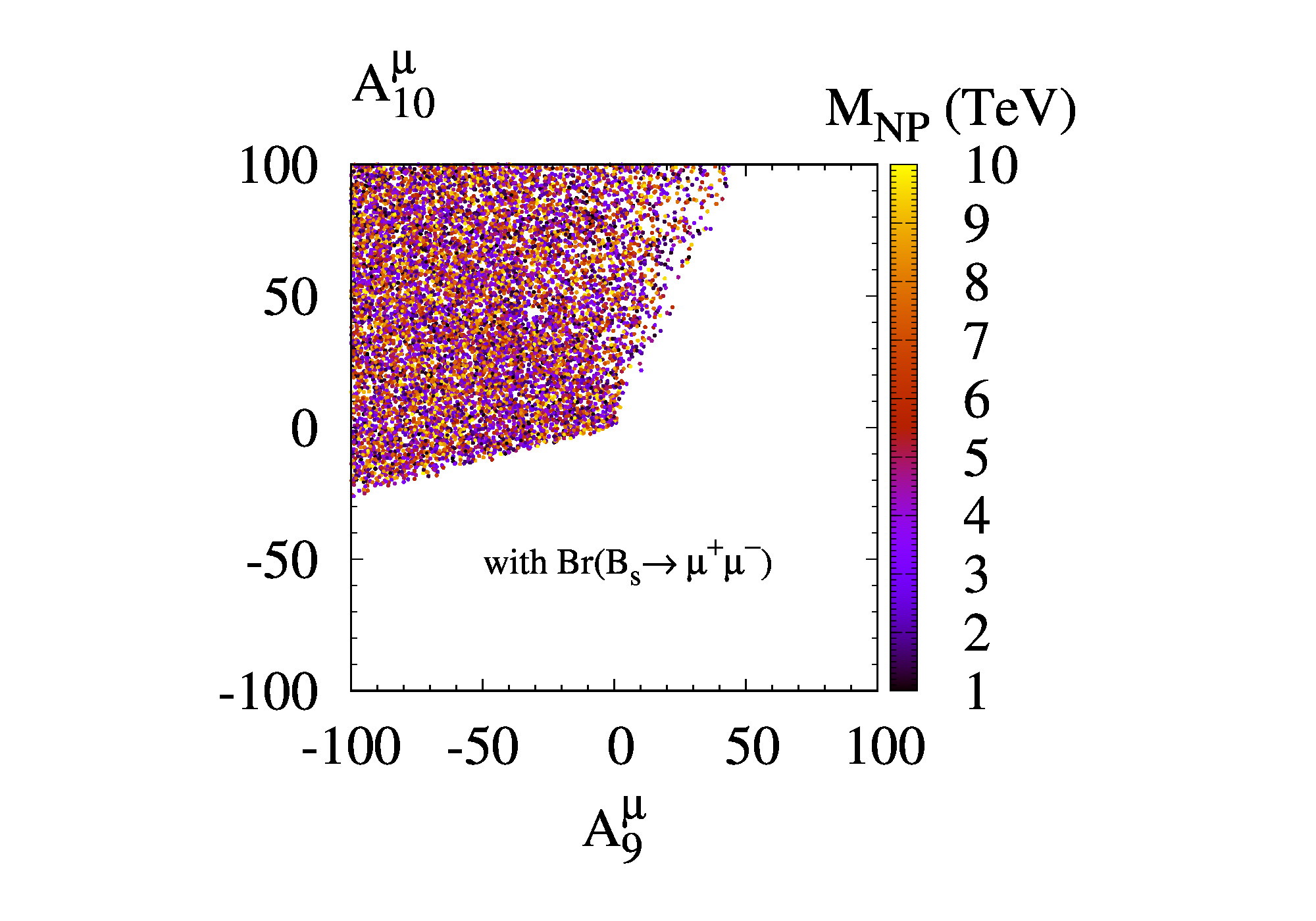}  \\
\hspace{-1cm} (a) & \hspace{-1cm} (b) \\
\hspace{-1cm}\includegraphics[scale=0.13]{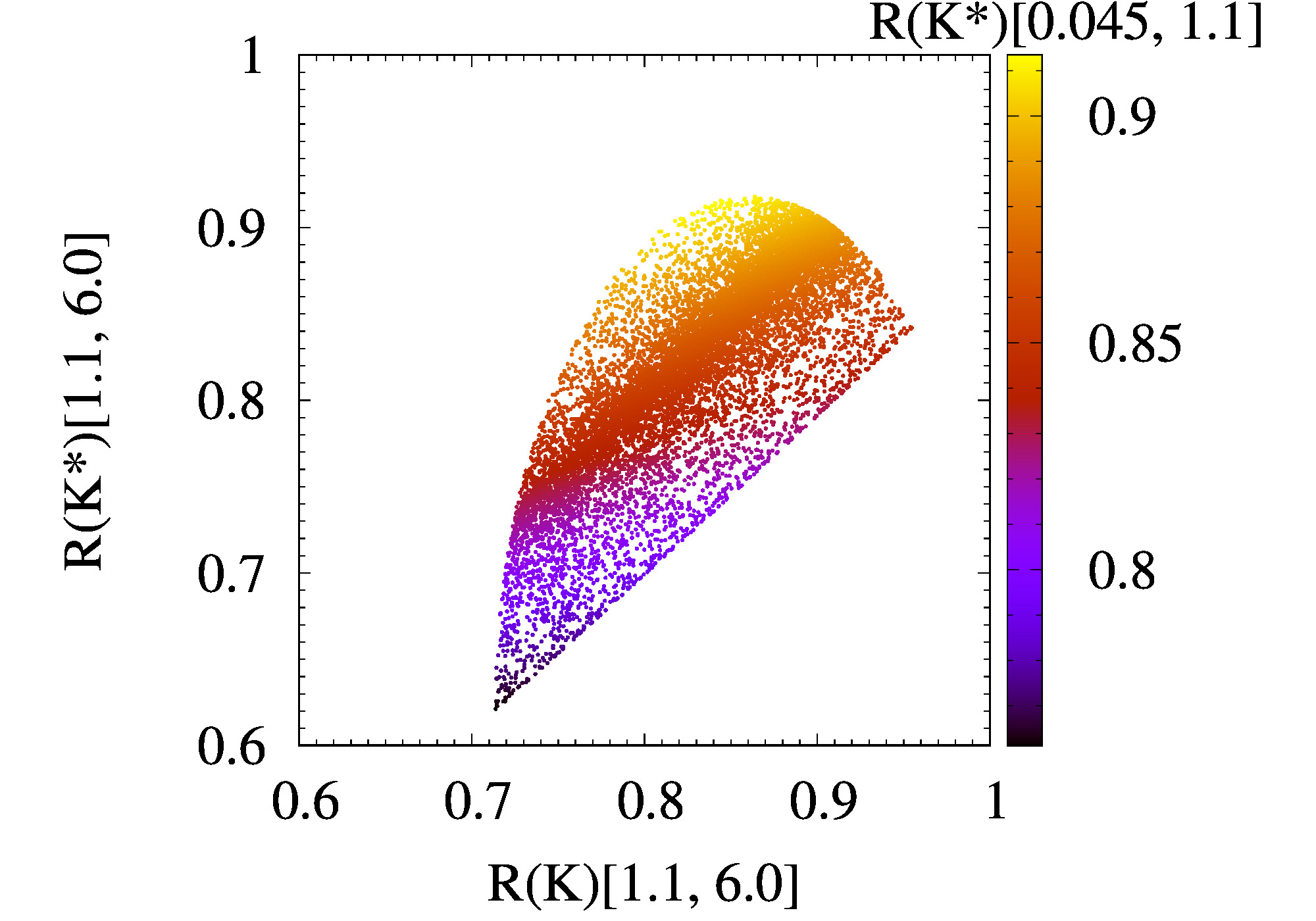} &
\hspace{-0cm}\includegraphics[scale=0.13]{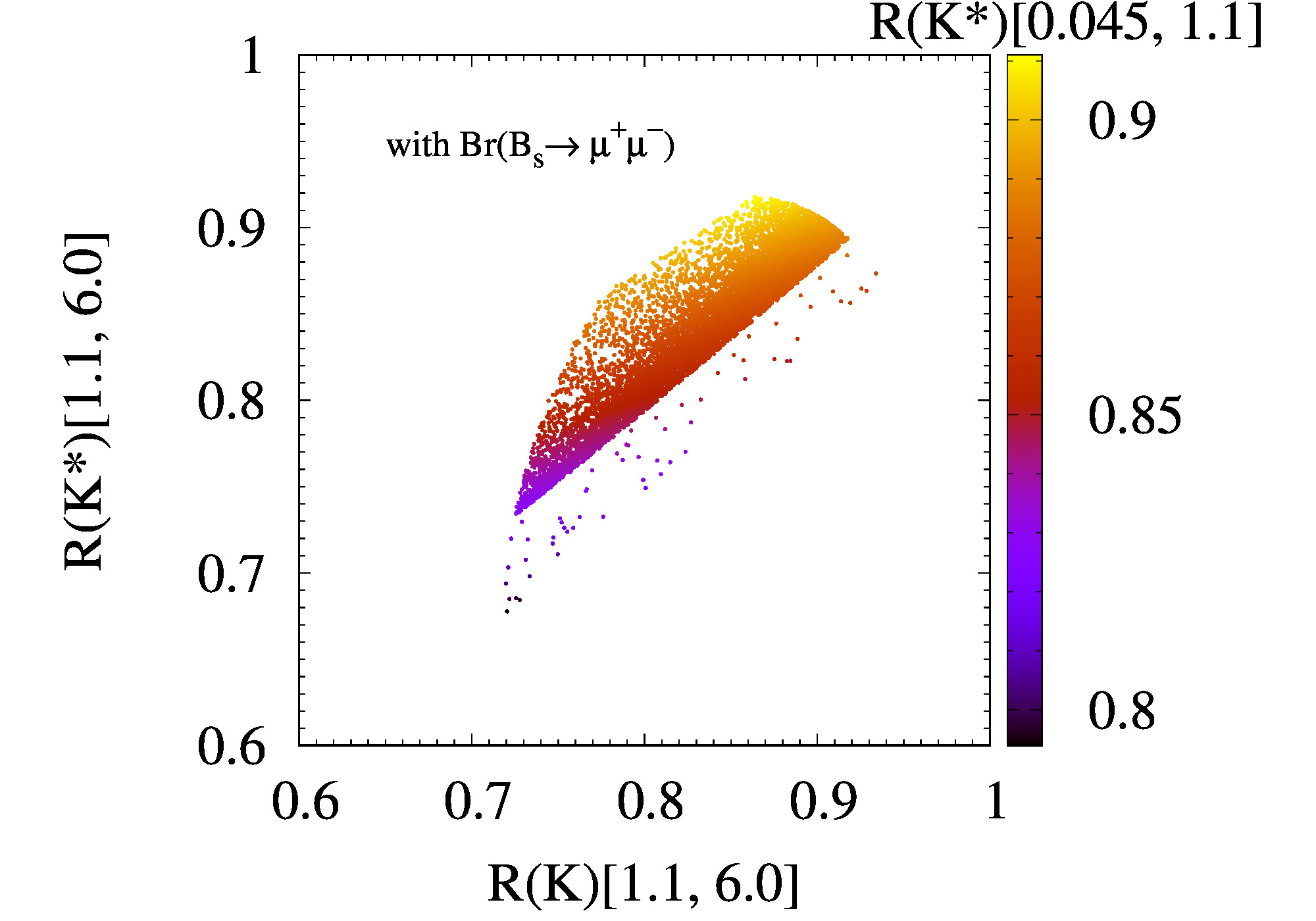}  \\
\hspace{-1cm} (c) & \hspace{-0cm} (d) 
\end{tabular}
\caption{\label{F3D} 
Allowed regions of (a) $A_{10}^\mu$ vs $A_9^\mu$ with respect to $M_\NP$
and (b) the subregion of (a) where $\Br(\Bs2mumu)$ is satisfied;
(c) $R(K^*)[1.1, 6.0]$ vs $R(K)[1.1,6.0]$ with respect to $R(K^*)[0.045, 1.1]$
and (d) the subregion of (c) where $\Br(\Bs2mumu)$ is satisfied,
at the $2\sigma$ level with $0\le\alpha\le 5$.}
\end{figure}

Figure \ref{F3D} is given for the allowed regions of 
$A_{10}^\mu$ vs $A_9^\mu$ with respect to $M_\NP$, and $R(\Ks)$.
Comparing Figs. \ref{F3D} (a) and (b), constraints from $\Bs2mumu$ reduce the allowed region
of $A_{10}^\mu$ more than that of $A_9^\mu$.
Much of the negative $A_{10}^\mu$ region is forbidden by $\Bs2mumu$, as discussed before.
And the distribution of $M_\NP$ is influenced little by $\Bs2mumu$, 
which is also clear in Fig.\ \ref{F1} (c).
%
As for $R(\Ks)$ in Figs.\ \ref{F3D} (c) and (d), parameters that enhance one of $R(\Ks)$s 
have a tendency to make other $R(\Ks)$s get larger.
Figure \ref{F3D} (d) shows that the constraint from $\Bs2mumu$ picks out the middle band-like 
region of Fig.\ \ref{F3D} (c).
In specific models, one should include other constraints like 
$(g-2)_\mu$, $B_s$-$\Bbar_s$ mixing, etc.
%
%
%
\par
Since $\alpha_\best=1.24$ unparticle-like scenario is favored more than other NP models where $\alpha=2$.
But for unparticles the scaling dimensions $d_S$ (for scalar) and $d_V$ (for vector) are 
$d_S=\alpha/2~, d_V=\alpha/2+1$ , 
and the unitarity condition puts bounds as $d_S\ge 1$ and $d_V\ge 3$.
If the unitary bound could be alleviated \cite{JPL2106} then $d_V$ could be $d_V\ge 1~(\alpha\ge 0)$,
so the vector unparticle might be a good candidate.
For the case of $\alpha=2$ where real particles are exchanged at the tree level,
we found that the minimum $\chi^2$ for the LQ model where
$-25\le A_{9,\LQ}(=-A_{10,\LQ})\le -2.5$ and $2\le M_\NP \le 4~(\TeV)$ \cite{Du2104}
is $\chi^2_{\min,\LQ}/\dof=4.39$.
As a comparison for $Z'$ model where 
$-27.2\le A_{9,Z'}(=-A_{10,Z'})\le -3.35$ and $3\le M_\NP\le 6~(\TeV)$ \cite{Davighi2105}, we got
$\chi^2_{\min,Z'}/\dof=7.75$.
At least for these windows of the model parameters one can say that LQ is better than $Z'$ to fit $R(\Ks)$.
As for sbottom exchange in SUSY of \cite{Altmannshofer2002} with 
$-30\le A_{9,{\rm SUSY}}(=-A_{10,{\rm SUSY}})\le 0$ and $1\le M_\NP\le 5~(\TeV)$,
we found $\chi^2_{\min,{\rm SUSY}}/\dof=4.30$.
The scan window for sbottom exchange is slightly wider than that of LQ or $Z'$, which results in better $\chi^2$.
For other models we could do similar comparisons.
\par
A comment is in order for the form-factor dependence.
Current analysis is based on the exponential function for the relevant form factors as in Eq.\ (\ref{expFF}).
In \cite{JPL2110}, outputs from different kind of form factors which have the pole structure
are compared.
The result is that general features are quite the same though some details are different.
It seems that the form-factor dependence in the numerators of $R(\Ks)$ is cancelled out by
the one in the denominators.
This kind of cancellation can be found in the hadronic uncertainties in the SM \cite{Geng1704}.
%
%
%
%
%
%
\par
Finally, we give comments on the new experimental data from the LHCb \cite{LHCb2212_52,LHCb2212_53}.
Compared to the previous results, the numbers get enhanced close to unity (and to the SM predictions) \cite{LHCb2212_53}
\begin{eqnarray}
R(K)[{\rm low}~ q^2] &=& 0.994^{+0.094}_{-0.087}~,~~~
R(K)[{\rm central}~ q^2] = 0.949^{+0.048}_{-0.047}~,\\
R(K^*)[{\rm low}~ q^2] &=& 0.927^{+0.099}_{-0.093}~,~~~
R(K^*)[{\rm central}~ q^2] = 1.027^{+0.077}_{-0.073}~.
\end{eqnarray}
The new data are far away from the previous one and if we include the above data the $\chi^2$ fitting
would get worse.
A preliminary analysis says that when including new data the $\chi^2_\min/\dof$ gets larger,
$\chi^2_\min/\dof =3.05$ as expected.
If we discard old data and keep only new data, then it would provide a strong constraint for NP.
%
\section{Conclusions}
In conclusion, we performed the $\chi^2$ fit to $R(\Ks)$ with a general parametrization of 
the NP effects for the Wilson coefficients by using the NP scale $M_\NP$, its power $\alpha$,
and the corresponding coefficients $A_{9,10}^\mu$.
In this framework the constraint from $\Bs2mumu$ is naturally incorporated with $R(\Ks)$.
This kind of parametrization for NP includes the leptoquark model, $Z'$ model, SUSY, and so on.
It is known that the relevant Wilson coefficients of $\sim\calO(1)$ is compatible with the data.
Our parametrization makes it easy to produce $\sim\calO(1)$ Wilson coefficients 
within natural ranges of the parameters.
In this analysis we considered the NP effects only in the muon sector.
Our results for the Wilson coefficients include the usual solution where 
$C_{9\NP}^\mu=-C_{10\NP}^\mu$.
We found that for the coefficients of $|A_{9,10}^\mu|$ up to $\sim\calO(2)$ with $\alpha=2$,
the NP scale $M_\NP$ is restricted to be $M_\NP\lesssim 5\TeV$.
The upper limit of $M_\NP$ gets higher for larger $|A_{9,10}^\mu|$.
Current analysis was based on the exponential function for the form factors of the matrix elements,
but it is expected that the form-factor dependence would be weak.
Our framework would work as a good starting point for specific models to attack the $R(\Ks)$ puzzle.

\end{document}